\documentclass[a4paper]{elsarticle}
\usepackage{graphicx}
\usepackage{epstopdf}
\usepackage{amsmath, amsthm, amssymb}
\usepackage{tikz}
\usepackage{pgflibraryarrows}
\usepackage{color}
\usepackage{booktabs}
\usepackage{arydshln}
\usepackage[margin=3.5cm]{geometry}
\usepackage{mathtools}
\usepackage{hyperref}
\usepackage[mathlines]{lineno}
\usepackage{multirow}

\usetikzlibrary{decorations.markings}
\usetikzlibrary{shapes.misc, positioning}
\definecolor{node_color}{gray}{0.925}

\newcommand{\btheta}{\boldsymbol{\theta}}

\DeclarePairedDelimiter{\norm}{\lVert}{\rVert}

\makeatletter
\def\ps@pprintTitle{%
 \let\@oddhead\@empty
 \let\@evenhead\@empty
 \def\@oddfoot{}%
 \let\@evenfoot\@oddfoot}
\makeatother

\begin{document}
\sloppy
\begin{frontmatter}

\title{High-dimensional structure learning of binary pairwise Markov networks: A comparative numerical study}

\author[lab1]{Johan Pensar\fnref{ca}}
\author[lab2,lab3]{Yingying Xu}
\author[lab2,lab3,lab5]{Santeri Puranen}
\author[lab2,lab3]{Maiju Pesonen}
\author[lab4]{Yoshiyuki Kabashima}
\author[lab1,lab5,lab6]{Jukka Corander}

\address[lab1]{Department of Mathematics and Statistics, Faculty of Science, University of Helsinki, Helsinki, Finland}
\address[lab2]{Department of Computer Science, Aalto University, Espoo, Finland}
\address[lab3]{Department of Computer Science, University of Helsinki, Helsinki, Finland}
\address[lab4]{Department of Mathematical and Computing Science, Tokyo Institute of Technology, Tokyo, Japan}
\address[lab5]{Department of Biostatistics, University of Oslo, Oslo, Norway}
\address[lab6]{Parasites and Microbes, Wellcome Sanger Institute, Cambridge, United Kingdom}
\fntext[ca]{Corresponding author: johan.pensar@helsinki.fi}

\begin{abstract}
Learning the undirected graph structure of a Markov network from data is a problem that has received a lot of attention during the last few decades. As a result of the general applicability of the model class, a myriad of methods have been developed in parallel in several research fields. Recently, as the size of the considered systems has increased, the focus of new methods has been shifted towards the high-dimensional domain. In particular, introduction of the pseudo-likelihood function has pushed the limits of score-based methods which were originally based on the likelihood function. At the same time, methods based on simple pairwise tests have been developed to meet the challenges arising from increasingly large data sets in computational biology. Apart from being applicable to high-dimensional problems, methods based on the pseudo-likelihood and pairwise tests are fundamentally very different. To compare the accuracy of the different types of methods, an extensive numerical study is performed on data generated by binary pairwise Markov networks. A parallelizable Gibbs sampler, based on restricted Boltzmann machines, is proposed as a tool to efficiently sample from sparse high-dimensional networks. The results of the study show that pairwise methods can be more accurate than pseudo-likelihood methods in settings often encountered in high-dimensional structure learning applications.
\end{abstract}


\end{frontmatter}

\section{Introduction}
Learning the dependency pattern over a large collection of variables is an important problem encountered in various fields of science \cite{Koller09}. This learning task can be cast as a sub-problem in the popular class of probabilistic graphical models. The ultimate goal of graphical models is to represent the joint distribution over a collection of variables by exploiting statements of conditional independence. The key feature of graphical models is to use a graph structure to compactly encode the dependence structure over the variables. In addition to being an essential part of graphical models, the graph structure provides a natural target for applications where the main goal is to gain an understanding in how variables within a system interact with each other. 

In this work, we will consider the problem of learning the undirected graph structure of pairwise Markov networks over binary variables. As a result of the general applicability of Markov networks, new and exciting structure learning methods have been developed in parallel in various fields including machine learning \cite{Hofling09}, statistics \cite{Ravikumar10}, statistical physics \cite{Ekeberg13,Ekeberg14}, and computational biology \cite{Feizi13,Butte00,Margolin06,Faith07}. In particular, as the size of the considered problems has increased, computational scalability of the algorithms has become more important. For example, the huge number of variables encountered in genome-wide sequencing studies in computational biology has steered the research towards methods based on simple pairwise tests \cite{Feizi13,Butte00,Margolin06,Faith07}. Although pairwise methods do not come with any general asymptotic guarantees that would hold in a general setting, they have been shown to perform well on various real-world applications. In contrast, the pseudo-likelihood methods, which have been developed in the more theoretical fields of statistics \cite{Ravikumar10} and statistical physics \cite{Ekeberg13,Ekeberg14}, are more elaborate and enjoy nice theoretical properties, such as consistency.

Although pairwise and pseudo-likelihood methods have been designed for the same underlying problem, they have rarely been compared against each other in a controlled setting. In particular, the finite-sample performance of the pseudo-likelihood methods in comparison to the pairwise methods has not been thoroughly examined. Therefore, in this work, we perform a numerical simulation study on synthetic models to compare a collection of methods from the two algorithmic families. The focus of the study is on high-dimensional sparse models, which are often encountered in real-world applications. To facilitate sampling from such models, we present a parallelizable and highly scalable Gibbs sampler based on sparse restricted Boltzmann machines.

We begin in Section \ref{sec:mn} by introducing binary pairwise Markov networks. After that, we continue by presenting the Gibbs sampler and introducing the structure learning problem. In Section \ref{sec:methods}, we present the methods included in the study. We discuss the main idea and the underlying assumptions behind each method. In Section \ref{sec:exp}, we go through the experimental setup and present the key findings. We conclude this work with a discussion in Section \ref{sec:discussion}.

\section{Pairwise Markov networks \label{sec:mn}}
A Markov network (or Markov random field) is an undirected graphical model that represents a joint distribution over a set of $d$ random variables $X=\{X_{1},\ldots,X_{d}\}$. In this work, we assume a positive distribution and we also restrict the variables to be binary with outcome space $\mathcal{X}_v=\{ 0,1\}$. We use a lowercase letter $x_{v}$ to denote that the corresponding variable has been assigned a specific value in $\mathcal{X}_v$. The dependence structure of a Markov network is encoded by an undirected graph $G=(V,E)$, where $V=\{ 1,\ldots ,d\}$ are nodes (or vertices), corresponding to the variables, and $E\subseteq V\times V$ are edges, representing direct dependencies between the variables. In particular, absence of edges implies statements of conditional independence, such that separation between nodes in the graph implies conditional independence between the corresponding variables (see e.g. \cite{Koller09} for more details). The conditional independence statements are the fundamental assumptions used by graphical models to decompose the joint distribution into smaller components. 

In this work, we will consider a special subclass of Markov network that only includes pairwise interactions. The (positive) joint distribution over the variables can then be parameterized as the log-linear model:
\[
p(x;\btheta)=\frac{1}{Z(\btheta)}\exp \left( \sum_{v\in V} x_v \theta_v + \sum_{\{v,v'\} \in E} x_v x_{v'} \theta_{vv'} \right),
\]
where $\btheta$ contains the model parameters, which consist of the node-specific bias parameters $\{\theta_v \}_{v\in V}$ and the edge-specific interaction parameters $\{\theta_{vv'} \}_{\{v,v'\}\in E}$ \cite{Whittaker90}. The function $Z(\btheta)$ is a normalizing constant known as the partition function:
\[
Z(\btheta)=\sum_{x\in\mathcal{X}}\exp \left( \sum_{v\in V} x_v \theta_v + \sum_{\{v,v'\} \in E} x_v x_{v'} \theta_{vv'} \right).
\]
For notational convenience, both $\theta_{vv'}$ and $\theta_{v'v}$ denote the same interaction parameter, unless otherwise stated. Binary pairwise Markov networks are equivalent to Ising models, which are a well-studied model class in the statistical physics community.

\subsection{Sampling}
Sampling from a non-chordal Markov network is an important and non-trivial problem. The common approach is to use MCMC methods, the most well-known being the sequential Gibbs sampler. In each iteration, the algorithm runs through the variables in a pre-specified order and updates the value of each variable by conditioning on the rest of the variables or, for a specified graph, the neighbors of the variable. In this work, we will use an alternative MCMC sampler based on sparse Restricted Boltzmann Machines (sRBM) with real-valued hidden variables. The approach is similar to method in \cite{Martens10}, where an RBM is used to sample fully-connected networks.

Let $G=(V,E)$ be the graph of a binary pairwise Markov network. Also, for the sake of the derivation in this section, consider the following matrix-based representation of the distribution:
\begin{equation}\label{eq:mndist_mat}
p(x;A,b)=\frac{1}{Z(A,b)}\exp \left(\frac{1}{2} x^{\top}Ax  + b^{\top}x \right),
\end{equation}
where $A$ is an interaction matrix, such that $A_{(v,v')}=\theta_{vv'} \text{ if } \{ v,v' \}\in E$ and 0 otherwise, and $b=(\theta_1,\ldots,\theta_d)$ is a vector containing the bias parameters.

For each $\{ v,v' \}\in E$, we introduce an auxiliary continuous variable, which we (with some abuse of notation) denote by $Y_{vv'}$. Thus, the number of auxiliary variables will be equal to the number of edges, which is denoted by $n_\text{edges}$. Now, the joint distribution over $(X,Y)$ will be defined in form of an sRBM, where the original variables $X$ are visible and the auxiliary variables $Y$ are hidden: 
\[
p(x,y; W,b)\propto \exp\left(y^{\top} W x -\frac{1}{2}y^{\top}y + b^{\top}x - \frac{1}{2}\text{diag}(W^{\top}W)^{\top}x\right),
\]
where $W$ is a sparse $n_\text{edges}$-by-$d$ edge weight matrix connecting $X$ and $Y$. Note that there are no direct interactions within $X$ or $Y$. Let $W_{(vv',:)}$ represent the row corresponding to variable $Y_{ vv'}$ and assume that $v<v'$. Each row $W_{(vv',:)}$ will contain two non-zero elements in columns $v$ and $v'$, which connect $X_v$ and $X_{v'}$ indirectly via $Y_{vv'}$. More specifically, to maintain the original edge-specific interactions between the variables in $X$ in the sRBM, the non-zero elements in $W$ are specified according to
\[
\begin{cases}
W_{(vv',v)} = \sqrt{|\theta_{vv'}|}\\
W_{(vv',v')} = \text{sign}(\theta_{vv'})\sqrt{|\theta_{vv'}|} 
\end{cases}
\text{ for all } \{v,v'\}\in E.
\]
%The non-zero elements in REF correspond to edge weights between x and y (see the example in Figure REF). 
%By rewriting equation REF in an alternative form,
%\[
%p(x,z)\propto \exp\left(\frac{1}{2}x^{\top}W^{\top}Wx -\frac{1}{2}(z-Wx)^{\top}(z-Wx) + f(x)\right),
%\]
%it follows that
%\begin{equation*}
%p(z\mid x)=\frac{ \exp\left(-\frac{1}{2}(z-Wx)^{\top}(z-Wx)\right)}{\int \exp\left(-\frac{1}{2}(z-Wx)^{\top}(z-Wx) \right) dz}.
%\end{equation*}
%From REF it is clear that $Z\mid X \sim \mathcal{N}(Wx,I)$ and consequently
%\[
%\int \exp\left(-\frac{1}{2}(z-Wx)^{\top}(z-Wx) \right) dz = (2\pi)^{n_e/2}
%\]
Following the same line of reasoning as \cite{Martens10}, the marginal distribution over $X$ can be shown to be
\begin{equation}\label{eq:xdist_rbm}
p(x;W,b) \propto \exp\left(\frac{1}{2}x^{\top}W^{\top}Wx + b^{\top}x - \frac{1}{2}\text{diag}(W^{\top}W)^{\top}x \right).
\end{equation}
As a result of how $W$ was constructed, we further have that
\begin{equation}\label{eq:xdist_eq}
\frac{1}{2} x^{\top}W^{\top}Wx = \frac{1}{2} x^{\top}Ax + \frac{1}{2} \text{diag}(W^{\top}W)^{\top}x,  
\end{equation}
which ultimately means that the marginal distribution \eqref{eq:xdist_rbm} is equivalent to \eqref{eq:mndist_mat}. In other words, we have constructed an sRBM for which the marginal distribution over the visible variables is identical to the joint distribution of the initial Markov network. 

\begin{figure}
\begin{center}
\begin{tikzpicture}
\node[at={(2,-1.2)}]{(a)};
\node[at={(10,-1.2)}]{(b)};
\tikzstyle{every node}=[draw,circle,fill=node_color,line width=0.3mm,inner sep=0,minimum size=20pt];
\node[at={(0,0)}](1){$X_1$};
\node[at={(2,0)}](2){$X_2$};
\node[at={(4,0)}](3){$X_3$};
\draw[line width=0.3mm] (1) -- (2);
\draw[line width=0.3mm] (2) -- (3);
\node[at={(8,0)}](1){$X_1$};
\node[at={(10,0)}](2){$X_2$};
\node[at={(12,0)}](3){$X_3$};
\tikzstyle{every node}=[draw,circle,line width=0.3mm,inner sep=0,minimum size=20pt];
\node[at={(9,1.5)}](11){$Y_{12}$};
\node[at={(11,1.5)}](22){$Y_{23}$};
\draw[line width=0.3mm] (1) -- (11);
\draw[line width=0.3mm] (2) -- (11);
\draw[line width=0.3mm] (2) -- (22);
\draw[line width=0.3mm] (3) -- (22);
\end{tikzpicture}
\end{center}
\caption{A small example illustrating the mapping between Markov networks and sRBMs. Figure (a) shows the structure of a Markov network over three variables. Figure (b) shows a corresponding sRBM with three visible variables (gray nodes) and two hidden variables (white nodes), one for each edge in the original graph.\label{fig:ex}}
\end{figure}
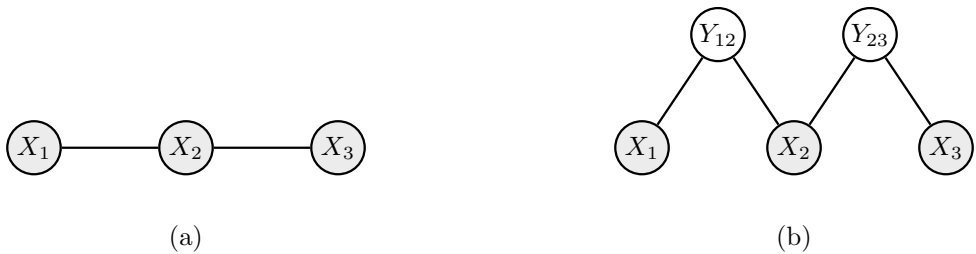

To give an example of the mapping between Markov networks and sRBMs, consider the Markov network structure in Figure \ref{fig:ex}(a) and the corresponding sRBM structure in Figure \ref{fig:ex}(b). For each edge in the Markov network, a hidden variable is added to the sRBM, connecting the visible variables indirectly. Moreover, for some interaction matrix
\[
A=
\begin{pmatrix}
0 & \theta_{12} & 0 \\
\theta_{12} & 0 & -\theta_{23} \\
0 & -\theta_{23} & 0 \\
\end{pmatrix},
\]
where $\theta_{12},\theta_{23}>0$, we have
\[
W=
\begin{pmatrix}
\sqrt{\theta_{12}} & \sqrt{\theta_{12}} & 0 \\
0 & \sqrt{\theta_{23}} & -\sqrt{\theta_{23}} \\
\end{pmatrix}
\hspace{0.2cm} \text{ and } \hspace{0.2cm}
W^{\top}W =
\begin{pmatrix}
\theta_{12} & \theta_{12} & 0 \\
\theta_{12} & \theta_{12}+\theta_{23} & -\theta_{23} \\
0 & -\theta_{23} & \theta_{23} \\
\end{pmatrix}.
\]
Note that the off-diagonal elements in $A$ and $W^{\top}W$ are identical, which is in line with equation \eqref{eq:xdist_eq}.

Clearly, the visible variables are mutually conditionally independent given the hidden variables, and vice versa, allowing for parallel sampling of individual variables. Similar to \cite{Martens10}, the conditional distributions of the hidden variables given the visible variables are normal: 
\[
Y_{vv'} \mid X=x \sim N(W_{(vv',:)}x,1) \hspace{0.25cm} \text{ for } \{v,v'\}\in E.
\]
Moreover, the conditional distributions of the visible variables given the hidden variables are given by
\[
p(X_v = x_v \mid Y=y) = \sigma\left((\theta_v - \frac{1}{2}(W^{\top}W)_{(v,v)})x_v + y^{\top}W_{(:,v)}\right) \hspace{0.25cm} \text{ for } v\in V, 
\]
where $\sigma()$ is the logistic sigmoid function and $W_{(:,v)}$ is the $v$:th column of $W$. Note that $W_{(vv',:)}$ contains only two non-zero elements and that $W_{(:,v)}$ contains as many non-zero elements as there are neighbors of $v$ in $G$. 

The mapping between Markov networks and sRBMs provides a straightforward technique for sampling from the joint distribution using a block-Gibbs sampler, which alternates between sampling the visible and hidden variables, and ultimately saving the configurations over the visible variables. Compared to a sequential Gibbs sampler, RBM samplers have been shown to mix slightly slower \cite{Martens10,Xu11} and they also require an additional step to sample the hidden variables. Still, from a computational perspective, the parallelization-friendly structure of an RBM sampler makes it a very attractive candidate for ultra-high-dimensional Markov network sampling using GPUs or other massively parallel environments.

\subsection{Structure learning}
Structure learning refers to the process of inferring the graph structure of a model from a set of data generated from that model. The data is assumed to contain $n$ complete i.i.d. joint observations which are represented by an $n$-by-$d$ data matrix $\mathbf{x}=(x^{(1)},\ldots,x^{(n)})$. One of the main challenges of structure learning is to distinguish between direct interactions (or edges) and indirect interactions mediated solely by multi-step paths over other nodes.

Structure learning methods have traditionally been divided into two categories; constraint-based and score-based. Constraint-based methods aim at recovering the model structure using a series of independence tests. Score-based methods formulate the structure learning problem as an optimization problem, which requires a score-function for evaluating how well a graph fits the data and an algorithm for optimizing the score function. 

\section{Structure learning methods \label{sec:methods}} 
The methods considered in this work have been developed in various fields spanning statistics, statistical mechanics and computational biology. The methods are from two conceptually very different algorithmic families: \emph{pseudo-likelihood methods}, which are score-based, and \emph{pairwise methods}, which are constraint-based using pairwise tests. Common to these methods is that they are applicable to high-dimensional problems and that they provide a way to rank edges by interaction strength. In this section we provide a brief description of the selected methods. For more details, we refer the reader to the original works.

\subsection{Pseudo-likelihood methods}
The (log-)likelihood function is probably the most important and well-known function in score-based model learning:
\[
l(\btheta;\mathbf{x})=\log p(\mathbf{x};\btheta)=\sum_{i=1}^{n} \log p(x^{(i)};\btheta). 
\]
In terms of structure learning, the underlying rationale is that $\theta_{vv'}=0$ iff $\{ v,v' \} \not\in E$. However, maximizing the likelihood of non-chordal Markov networks is computationally intractable, since computing the partition function $Z(\btheta)$ requires summing over all joint outcomes. Consequently, likelihood-based techniques are limited to small $d$ problems. 

As a computationally more convenient alternative, Besag \cite{Besag75} introduced the pseudo-likelihood, which is an approximation of the likelihood consisting of node-wise conditional likelihoods:
\begin{equation}\label{eq:cpd}
pl(\btheta;\mathbf{x})=\sum_{v=1}^{d} \log p(\mathbf{x}_v \mid \mathbf{x}_{-v};\btheta_v)=\sum_{v=1}^{d} \sum_{i=1}^{n} \log p(x_{v}^{(i)} \mid x_{-v}^{(i)};\btheta_v),
\end{equation}
where $-v=V\setminus v$ and $\btheta_v=(\theta_{v1},\ldots,\theta_{vd})$ is the parameter vector associated with $v$ such that $\theta_{vv}=\theta_v$. Initially, optimizing the pseudo-likelihood may seem like a cumbersome task. However, the main computational advantage, compared to the likelihood, is that the intractable partition function $Z(\btheta)$ cancels out in the conditional distributions. More specifically, the conditional distributions of a node $v$ is given by
\[
p(x_{v} \mid x_{-v};\btheta_v)=\frac{1}{Z_v(\btheta_v)}\exp \left( x_v \theta_v + \sum_{v' \in -v} x_v x_{v'} \theta_{vv'} \right),
\]
where 
\[
Z_v(\btheta_v)=1 + \exp \left( \theta_v + \sum_{v' \in -v} x_{v'} \theta_{vv'} \right)
\]
is a local normalizing constant depending on $\btheta_v$. %Note that for a given graph $G=(V,E)$, the set $-v$ is replaced by the (usually) considerably smaller neighbor set $nb(v)=\{v'\in V: \{v,v'\}\in E\}$ in the above formula, since $\theta_{vv'}=0$ iff $v'\not\in nb(v)$. 
In addition to being computationally tractable, the pseudo-likelihood function is concave and hence has no local maxima. Also, the maximum pseudo-likelihood estimator is consistent in the sense that the parameter values will approach the true values as $n\to\infty$ \cite{Hyvarinen06}. 

Moreover, examining \eqref{eq:cpd} closer, we see that the conditional distribution takes the form of a logistic regression. From now on, assume that $\theta_{vv'}$ and $\theta_{v'v}$ represent distinct parameters. Under the equality constraint $\theta_{vv'} = \theta_{v'v}$, maximizing the pseudo-likelihood can then be thought of as maximizing a sum of coupled logistic regression problems:
%\begin{equation*}
%\begin{aligned}
%&\underset{\btheta_v}{\arg\max} \ \sum_{v=1}^{d} \log p(\mathbf{x}_v \mid \mathbf{x}_{-v};\btheta_v) \\
%&\text{subject to } \hspace{0.3cm} \theta_{vv'} = \theta_{v'v}
%\end{aligned}
%\end{equation*}
\[
\underset{\btheta}{\arg\max} \ \sum_{v=1}^{d} ( (\mathbf{x} \btheta_{v})^{\top}\mathbf{x}_v - n\log Z_v(\btheta_v)).
\]
Following \cite{Ekeberg14}, we refer to this technique as the symmetric pseudo-likelihood approach. Symmetric pseudo-likelihood methods for structure learning in Markov networks have been developed in \cite{Hofling09,Ekeberg13}. 

As $d$ grows, optimizing the symmetric pseudo-likelihood eventually becomes too demanding. However, using the above formulation, it is straightforward to apply an even more efficient approach, obtained by simply removing the parameter equality constraint. This results in $d$ decoupled logistic regression problems:
%\begin{equation*}
%\underset{\btheta_v}{\arg\max} \ \log p(\mathbf{x}_v \mid \mathbf{x}_{-v};\btheta_v) \hspace{1cm}v=1,\ldots,d.
%\end{equation*}
\[
\underset{\btheta_v}{\arg\max} \ (\mathbf{x} \btheta_{v})^{\top}\mathbf{x}_v - n\log Z_v(\btheta_v) \hspace{1cm}v=1,\ldots,d.
\]
This approach, which is referred to as asymmetric pseudo-likelihood or regression-based structure learning, is very convenient from a computational perspective, since the subproblems can be solved independently of each other on multiple cores. During optimization, the asymmetric approach considers $\theta_{vv'}$ and $\theta_{v'v}$ as distinct parameters and will thus (in practice) give out two different estimates for each interaction parameter. Consequently, the solution requires some form of post-processing to be consistent with a Markov network. 

In order to avoid overfitting and ill-posedness when maximizing the pseudo-likelihood, it is in general necessary to impose some form of regularization on the model parameters. Next, we will present two asymmetric pseudo-likelihood methods, which are characterized by their type of regularization.

\subsubsection{Asymmetric pseudo-likelihood with $L_1$-regularization (plmL1)}
$L_1$-regularization is a popular technique for estimating sparse models from data, since the $L_1$-penalty will drive non-informative parameters to zero. Consequently, the $L_1$-norm has been considered by several authors as a means to perform structure learning of Markov networks \cite{Lee06,Hofling09,Ravikumar10}. In particular, \cite{Ravikumar10} introduced the popular asymmetric pseudo-likelihood with $L_1$-regularization:
%\begin{equation*}
%\underset{\btheta_v}{\arg\max} \ \log p(\mathbf{x}_v \mid \mathbf{x}_{-v};\btheta_v) - \lambda \norm{\btheta_v}_1 \hspace{1cm}v=1,\ldots,d.
%\end{equation*}
\[
\underset{\btheta_v}{\arg\max} \ (\mathbf{x} \btheta_{v})^{\top}\mathbf{x}_v - n\log Z_v(\btheta_v) - \lambda \norm{\btheta_v}_1 \hspace{1cm}v=1,\ldots,d,
\]
for which statistical guarantees for consistent graph estimation were proven under certain coherence conditions imposed on the Fisher information matrix. The method is a member of the well-known family of Lasso algorithms, a technique that was pioneered by \cite{Meinshausen06} for covariance selection for Gaussian graphical models. Although the $L_1$-norm is not differentiable, the problem is still convex and there exist a wide variety of methods for solving high-dimensional $L_1$-regularization problems.

Since $L_1$-regularization will actually drive parameters to zero, it is straightforward to infer the graph from the estimated parameters, since a non-zero parameter implies existence of an edge. However, we may obviously come across situations where $|\theta_{vv'}|>0$ while $|\theta_{v'v}|=0$. The common technique for dealing with the asymmetry issue is to add an edge $\{v,v'\}$ if 
\[
\min \left(|\theta_{vv'}|,|\theta_{v'v}|\right)>0,
\]
or alternatively, if
\[
\max \left(|\theta_{vv'}|,|\theta_{v'v}|\right)>0.
\]
The above criteria, which are also known as the $\wedge$ and $\vee$ rules \cite{Meinshausen06}, have shown similar accuracy in previous numerical experiments \cite{Hofling09,Barber15}.

The strength of the sparsity-inducing regularization term is adjusted by tuning the penalty weight $\lambda$, such that the degree of sparsity increases with a larger penalty weight. By moving across an appropriately chosen range of decreasing $\lambda$-values, one obtains a so-called regularization path, along which the number of non-zero interaction parameters increases from $0$ to $d-1$. In other words, the regularization path offers a straightforward way for ranking the edges using the penalty weight. 

\subsubsection{Asymmetric pseudo-likelihood with $L_2$-regularization (plmDCA)}
$L_2$-regularization is one of the most common techniques for dealing with ill-posed problems in statistics and machine learning, with examples being linear regression \cite{Hoerl70} and classification with logistic regression or support vector machines \cite{Fan08}. Ekeberg et al. \cite{Ekeberg13} introduced the idea of combining the symmetric pseudo-likelihood for categorical data with $L_2$-regularization. The original method was later replaced with an asymmetric version \cite{Ekeberg14}:
%\begin{equation*}
%\underset{\btheta_v}{\arg\max} \ \log p(\mathbf{x}_v \mid \mathbf{x}_{-v};\btheta_v) + \lambda \norm{\btheta_v}_2^2 \hspace{1cm}v=1,\ldots,d.
%\end{equation*}
\[
\underset{\btheta_v}{\arg\max} \ (\mathbf{x} \btheta_{v})^{\top}\mathbf{x}_v - n\log Z_v(\btheta_v) - \lambda \norm{\btheta_v}_2^2 \hspace{1cm}v=1,\ldots,d,
\]
which was shown to give similar results as the original symmetric version. The method was named plmDCA, since it was originally developed for direct coupling analysis (DCA), which is the task of identifying direct couplings, or contacts, between amino-acid positions in protein chains from multiple sequence alignments. PlmDCA is currently one of the leading methods for performing DCA \cite{deOliveira16}. Being differentiable, the $L_2$-norm is a convenient penalty function from the optimization point of view.  

In contrast to the $L_1$-penalty, $L_2$-regularization does not force parameters to zero. Therefore, the regression parameters are turned into an interaction score. The interaction strength of an edge $\{v,v' \}$ is calculated as follows. First, the estimated interaction parameters involving $v$ and $v'$ are collected into two coupling matrices; one is the result from regressing $v$ on $-v$ and the other from regressing $v'$ on $-v'$. After adjusting the two coupling matrices by the Ising gauge, they are averaged into a single coupling matrix, of which the Frobenius norm is calculated to give an edge-specific interaction score. Finally, the interaction scores are adjusted by an average-product correction step, which was originally introduced to suppress certain effects related to multiple sequence alignments.

\subsection{Pairwise methods\label{sec:mi}}
In computational biology, there are a wide range of applications that can be formulated as structure learning problems, one of the most well-known being gene expression networks. In general, the data sets are high-dimensional and of type $n < d$. To tackle the high-dimensional aspect of such data sets, an array of more or less heuristic methods have been developed. Although often lacking any theoretical guarantees that would hold in a general setting, methods of this kind have led to several successful applications.

In this work, we will consider methods that are based on a simple pairwise score, which measures the marginal dependence between two variables. As our pairwise score, we will use mutual information, which is one of the most well-known measures of marginal (or mutual) dependence. The mutual information (MI) between two discrete variables, indexed by $v$ and $v'$, is defined by
\[
MI(v,v')=\sum_{x_v\in \mathcal{X}_v} \sum_{x_{v'}\in \mathcal{X}_{v'}} p(x_v,x_{v'}) \log\left( \frac{p(x_v,x_{v'})}{p(x_v)p(x_{v'})} \right).
\]
In practice, we do not have access to the actual joint distribution $p(X_v,X_{v'})$, which has to be estimated from data. Rather than using the maximum likelihood estimate, we use a smoothing procedure where we add a pseudo-count of $1/4$ to each joint outcome, giving a total pseudo-count of $1$. Thus, our estimates are calculated according to
\[
\hat p(x_v,x_{v'})=\frac{n(x_v,x_{v'})+1/4}{n+1}\hspace{0.35cm} \text{ and }\hspace{0.35cm} \hat p(x_v) =\sum_{x_{v'}\in \mathcal{X}_{v'}} \hat p(x_v,x_{v'}),
\]
where $n(x_v,x_{v'})$ represents the count of the corresponding outcome. From a Bayesian perspective, the estimator can be interpreted as the posterior mean, using a symmetric Dirichlet prior with hyperparameter $1/4$.

The most basic MI-based approach is to the calculate the mutual information for each pair of variables and then rank the strength of the edges according to their MI values (cf. Relevance networks \cite{Butte00}). There is an inherent flaw with this approach, since a pairwise test on its own cannot distinguish between direct and indirect dependencies. Consequently, a variety of post-processing techniques have been designed for dealing with this particular issue.

\subsubsection{Context likelihood of relatedness (CLR) }

The CLR algorithm was introduced as an approach to remove false positives in gene regulatory networks through a background correction step \cite{Faith07}. After computing the MI between all pairs of nodes, CLR calculates an adjusted interaction score that takes the background distribution of MI values into account. The background distribution of $MI(v,v')$ is approximated as a bivariate normal distribution estimated from the MI values involving the nodes $v$ and $v'$. More specifically, let $z_{v}(v,v')$ and $z_{v'}(v,v')$ be the z-scores of $MI(v,v')$ computed with respect to the MI values involving $v$ and $v'$, respectively. The background adjusted interaction score between $v$ and $v'$ is then calculated from the z-values according to
\[
f(v,v') = \sqrt{z_{v}(v,v')^2 + z_{v'}(v,v')^2}.
\] 
Before calculating the adjusted interaction score, negative z-values are set to zero. As a result, most weight will be given to interactions that stand significantly above the marginal background distributions of the considered variables.

\subsubsection{An algorithm for the reconstruction of gene regulatory networks (ARACNE)}\label{sec:aracne}
ARACNE was proposed as a means for disentangling direct from indirect dependencies when learning gene regulatory networks \cite{Margolin06}. The method is based on a notion referred to as the \emph{data processing inequality (DPI)}. The DPI states that if two nodes $v$ and $v'$ only interact indirectly through a third node $v''$, then 
\[
MI(v,v') \leq \min \left( MI(v,v''),MI(v',v'')\right).
\]
In other words, information between $v$ and $v'$ is in general lost (but never gained) when transmitted through $v''$.

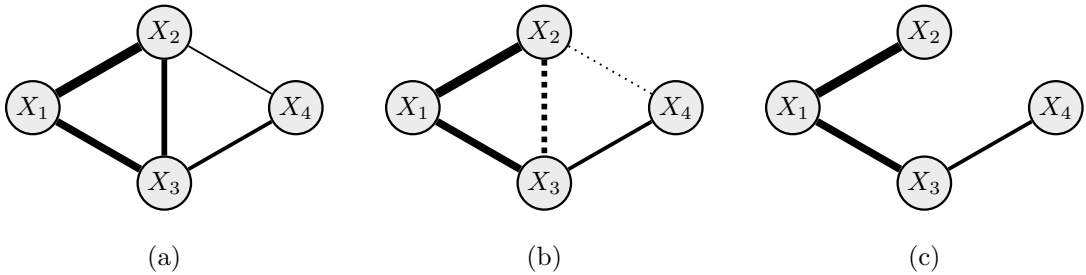
\begin{figure}
\begin{center}
\begin{tikzpicture}
\node[at={(1.73,-2)}]{(a)};
\node[at={(6.73,-2)}]{(b)};
\node[at={(11.73,-2)}]{(c)};
\tikzstyle{every node}=[draw,circle,fill=node_color,line width=0.3mm,inner sep=0,minimum size=20pt];
\node[at={(0,0)}](1){$X_1$};
\node[at={(1.73,1)}](2){$X_2$};
\node[at={(1.73,-1)}](3){$X_3$};
\node[at={(3.46,0)}](4){$X_4$};
\draw[line width=1.25mm] (1) -- (2);
\draw[line width=1mm] (1) -- (3);
\draw[line width=0.75mm] (2) -- (3);
\draw[line width=0.25mm] (2) -- (4);
\draw[line width=0.5mm] (3) -- (4);

\node[at={(5,0)}](1){$X_1$};
\node[at={(6.73,1)}](2){$X_2$};
\node[at={(6.73,-1)}](3){$X_3$};
\node[at={(8.46,0)}](4){$X_4$};
\draw[line width=1.25mm] (1) -- (2);
\draw[line width=1mm] (1) -- (3);
\draw[line width=0.75mm,dotted] (2) -- (3);
\draw[line width=0.25mm,dotted] (2) -- (4);
\draw[line width=0.5mm] (3) -- (4);

\node[at={(10,0)}](1){$X_1$};
\node[at={(11.73,1)}](2){$X_2$};
\node[at={(11.73,-1)}](3){$X_3$};
\node[at={(13.46,0)}](4){$X_4$};
\draw[line width=1.25mm] (1) -- (2);
\draw[line width=1mm] (1) -- (3);
\draw[line width=0.5mm] (3) -- (4);
\end{tikzpicture}
\end{center}
\caption{A small example illustrating the ARACNE filtering step: (a) an initial graph where the magnitude of the MI values are visualized by the thickness of the edges, (b) edge $2-3$ is removed, since it is weaker than both $1-2$ and $1-3$, and edge $2-4$ is removed, since it is weaker than both $2-3$ and $3-4$, (c) final graph after running the ARACNE filtering step. \label{fig:aracne_step}}
\end{figure}

ARACNE starts from a graph containing an edge for each non-zero MI value. The algorithm then proceeds by examining each triplet of mutually connected nodes. For each such triplet, the edge associated with the lowest MI value is removed if the difference to the second largest MI value in the triplet is larger than a user-specified tolerance threshold. The triplets are examined independently of each other, meaning that an edge that has been removed with respect to one triplet is still taken into consideration when examining another triplet containing that particular edge, see Figure \ref{fig:aracne_step} for an example. The output of the algorithm is therefore not affected by the order in which the triplets are examined.

ARACNE is better suited for certain types of graph topologies. In particular, for a Markov network with a tree-based dependence structure, ARACNE will recover the correct graph for a large enough sample. On the other hand, using a zero tolerance, loops over three nodes cannot be discovered, since the weakest edge in such loops will always be removed (except in the degenerate case). In general, ARACNE is well-suited for networks with locally tree-like structures in which the shortest path dominates the information exchange between two nodes.

\subsubsection{Network deconvolution (ND)}
The ND algorithm was introduced as a general purpose algorithm for filtering out the indirect effects from an observed correlation matrix containing both direct and indirect effects \cite{Feizi13}. By formulating the problem as the inverse of network deconvolution, the algorithm attempts to remove the indirect effects by using eigendecomposition and a result for geometric infinite series of matrices. The fundamental assumption behind the algorithm is that indirect dependencies can be approximated by products of direct dependencies such that the observed marginal dependencies are the sum of direct and indirect dependencies. Although the assumption does not hold in general, the method was shown to perform very well on a collection of real-world problems \cite{Feizi13}. 

Let $G_{\text{dir}}$ and $G_{\text{indir}}$ be symmetric matrices containing the direct and indirect effects of a Markov network. Also, let $G_{\text{obs}}$ be a matrix of observed marginal dependencies. Then ND assumes that
\[
 G_{\text{obs}} =  G_{\text{dir}} +  G_{\text{indir}} = G_{\text{dir}} +  (G_{\text{dir}}^2 +  G_{\text{dir}}^3 + \ldots).
\]
The goal is to recover the sparse $G_{\text{dir}}$ matrix from the observed $G_{\text{obs}}$. Under the assumption that the maximum absolute eigenvalue of $G_{\text{dir}}$ is strictly smaller than one, the infinite sum can be expressed as $G_{\text{obs}} =  G_{\text{dir}}(I + G_{\text{dir}})^{-1}$, which gives the closed form solution
\[
G_{\text{dir}} =  G_{\text{obs}}(I + G_{\text{obs}})^{-1}.
\]
In practice, the observed dependency matrix $G_{\text{obs}}$ is linearly scaled in order to make the absolute eigenvalues of $G_{\text{dir}}$ smaller than one. The above inverse operation is efficiently computed using the eigendecompositions of $G_{\text{obs}}$ and $G_{\text{dir}}$.
%\[
%G_{\text{obs}} = U\Sigma_{\text{obs}}U^{-1}\hspace{0.25cm}  \text{and} \hspace{0.25cm}  G_{\text{dir}} = U\Sigma_{\text{dir}}U^{-1},
%\]
%where $\lambda_{v}^{\text{dir}} = \lambda_{v}^{\text{obs}}/(1+\lambda_{v}^{\text{obs}})$. 

\section{Numerical experiments \label{sec:exp}}
To empirically compare the accuracy of the methods discussed in the previous section, we performed an extensive numerical simulation study. In addition to the presented methods, we included a plain MI-based method as a baseline. The purpose of the simulation was to investigate how the methods perform in practice in a variety of different scenarios, which have not been tailored for any particular method. In particular, we wanted to examine the relationship between the model size $d$ and the sample size $n$. For more details about the implementation of the methods, see \ref{app:impl}.

\subsection{Experimental setup}
To test if the methods are particularly ill- or well-suited for a specific type of network structure, we selected three standard network types with different structural characteristics:
\begin{itemize}
\item \emph{Grid network:} A fixed four-nearest neighbour grid graph (also known as a lattice graph).
\item \emph{Scale-free network:} A random scale-free network generated by the Barab{\'{a}}si-Albert model \cite{Barabasi99}. The algorithm starts from an initial connected network over $m_0$ nodes and successively adds new nodes such that each new node is connected to $m$ existing nodes. The algorithm uses a preferential attachment mechanism in which the probability of a node receiving an edge during the generation process is proportional to the number of edges already connected to that node.
\item \emph{Small-world network:} A random small-world network generated by the Watts-Strogatz model \cite{Watts98}. The algorithm starts by constructing a regular ring lattice in which each node is connected to its $k$ closest neighbors. After the initial step, the algorithm iterates through each existing edge and randomly rewires it with a probability $p$. 
\end{itemize}
For each network type, we generated 50 structures over 40, 200 and 1000 nodes (see \ref{app:impl} for specific model parameters). Various structural properties of the generated networks are listed in Table~\ref{tab:network_prop}. All of the generated networks are sparse, with the average number of neighbours varying between 3.35 (grid, $d=40$) and 4.20 (scale-free, $d=1000$). In contrast to the highly regular fixed grid networks, the algorithms behind the randomly generated scale-free and small-world networks are designed to capture certain structural properties observed in many real-world networks. More specifically, the scale-free model generates hub nodes through its preferential attachment mechanism, resulting in large maximum neighbourhoods. In contrast, the small-world model produces high levels of clustering, which is reflected by the average clustering coefficient in Table~\ref{tab:network_prop}.

%\begin{figure}
%\begin{center}
%\includegraphics[width=\textwidth]{figures/sample_graphs_d40.pdf}
%\end{center}
%\caption{Sample graphs ($d=40$) of the network types in the experiment.\label{fig:sample_graphs}}
%\end{figure}

\setlength{\tabcolsep}{6pt}
\begin{table}
\begin{center}
\small
\begin{tabular}{l l c c c}
\toprule
Network property & $d$ & Grid network & Scale-free network & Small-world network  \\
\midrule
& 40 & $3.35 \pm 0$ & $4.00 \pm 0.09$  & $4 \pm 0$ \\ 
Avg. neigbourhood size & 200 & $3.70 \pm 0$ & $4.15 \pm 0.04$ & $4 \pm 0$ \\ 
& 1000 & $3.91 \pm 0$ & $4.20 \pm 0.02$ & $4 \pm 0$ \\ 
\\
& 40 & $4 \pm 0$ & $10.16 \pm 1.17$ & $6.24 \pm 0.69$ \\ 
Max. neigbourhood size & 200 & $4 \pm 0$ & $14.40 \pm 1.60$ & $6.84 \pm 0.58$ \\ 
& 1000 & $4 \pm 0$ & $19.40 \pm 2.08$ & $7.82 \pm 0.60$ \\ 
\\
& 40 & $0 \pm 0$ & $0.11 \pm 0.03$ & $0.27 \pm 0.05$ \\ 
Avg. clustering coefficient & 200 & $0 \pm 0$ & $0.03 \pm 0.01$  & $0.23 \pm 0.02$ \\ 
& 1000 & $0 \pm 0$ & $0.01 \pm 0.00$ & $0.23 \pm 0.01$ \\ 
\bottomrule 
\end{tabular}
\end{center}
\caption{Various structural characteristics of the generated network structures ($\text{mean} \pm \text{sd}$). The average/maximum neighbourhood size is the average/maximum number of nodes connected to a node. The average clustering coefficient measures the average connectedness between neighbours of a node. \label{tab:network_prop}}
\end{table}

For each graph structure, the strength of the associated model parameters were drawn randomly according to
\[
| \theta_v | \sim \text{Uniform}(0,1)\hspace{0.25cm} \text{and}\hspace{0.25cm} | \theta_{vv'} | \sim \text{Uniform}(1,2),
\]
and the sign of a parameter was determined by a $\text{Bernoulli}(0.5)$ trial. From each model, we generated a data set of size 200, 1000 and 5000 using the sRBM sampler. In total, each method was thus applied on 1350 datasets covering $n/d$-ratios from $1/5$ up to $125$. 

\subsection{Method evaluation}
As a basis for evaluating the performance of the methods, we used \emph{precision} and \emph{recall}:
\[
\text{\emph{precision}} = \frac{TP}{TP + FP} \hspace{0.5cm}\text{and}\hspace{0.5cm} \text{\emph{recall}}=\frac{TP}{TP+FN}.
\]
Precision is the fraction of true edges among the included edges, while recall is the fraction of included true edges among the existing true edges. 

Since all methods provide a way to rank the edges according to their estimated interaction strength, we obtain a complete range of precision-recall pairs. Consequently, as our first evaluation measure, we consider the complete precision-recall curve summarized by the Area Under the Curve (AUC). The AUC thus provides a measure of the overall accuracy of the methods across the complete range of recall. 

As our second accuracy measure, we use the maximum recall reached by a method under a precision of 0.90 (denoted by $RC_{0.90}$), or equivalently, under a false discovery rate of 0.10. In many applications, it is critical to maintain a high precision for the included edges, since each potential edge needs to be verified manually, or in some cases even experimentally. A high value on $RC_{0.90}$ implies that a method is able to recover a high fraction of the true edges while keeping down the number false positives, in this case, only 1 out of 10 among the discovered edges is a false positive.  

%See Figure \ref{fig:ex_graphs} for sample graphs of each network type. 
%\begin{figure}
%\begin{center}
%\includegraphics[trim = 0mm 0mm 0mm 0mm, clip,width=1\textwidth]{figures/graphs.pdf}
%\end{center}
%\vspace{-1cm}
%\caption{Sample graphs ($d=100$) of the network types in the experiment.\label{fig:ex_graphs}}
%\end{figure}

\subsection{Results}
\begin{figure}
\begin{center}
\includegraphics[trim={1.5cm 1cm 1.5cm 3cm},clip,width=\textwidth]{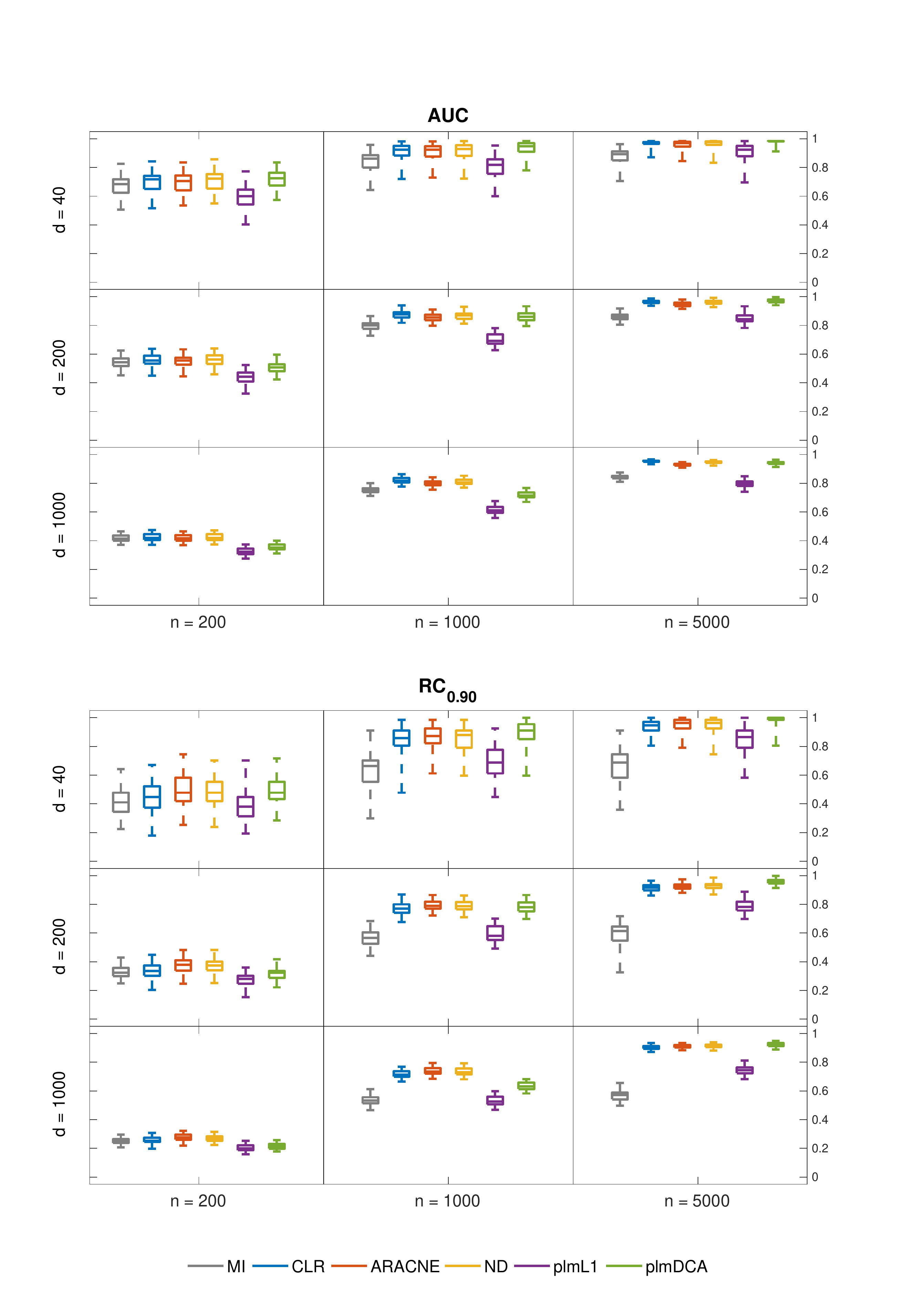}
\end{center}
\caption{Grid network: Area under the precision-recall curve (AUC) and maximum recall for a precision of 0.90 ($RC_{0.90}$).\label{fig:res_grid}}
\end{figure}

\begin{figure}
\begin{center}
\includegraphics[trim={1.5cm 1cm 1.5cm 3cm},clip,width=\textwidth]{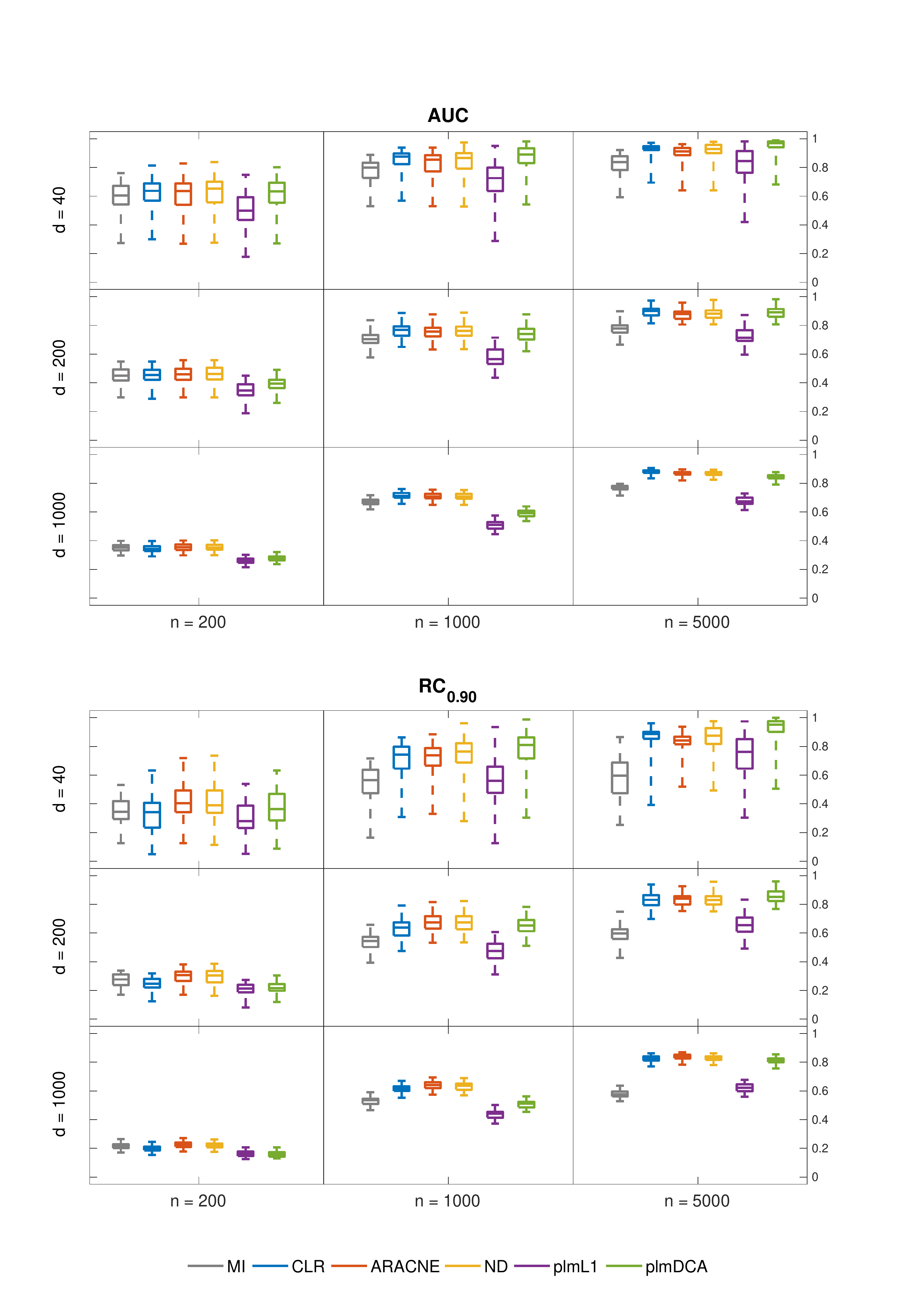}
\end{center}
\caption{Scale-free network: Area under the precision-recall curve (AUC) and maximum recall for a precision of 0.90 ($RC_{0.90}$).\label{fig:res_ba}}
\end{figure}

\begin{figure}
\begin{center}
\includegraphics[trim={1.5cm 1cm 1.5cm 3cm},clip,width=\textwidth]{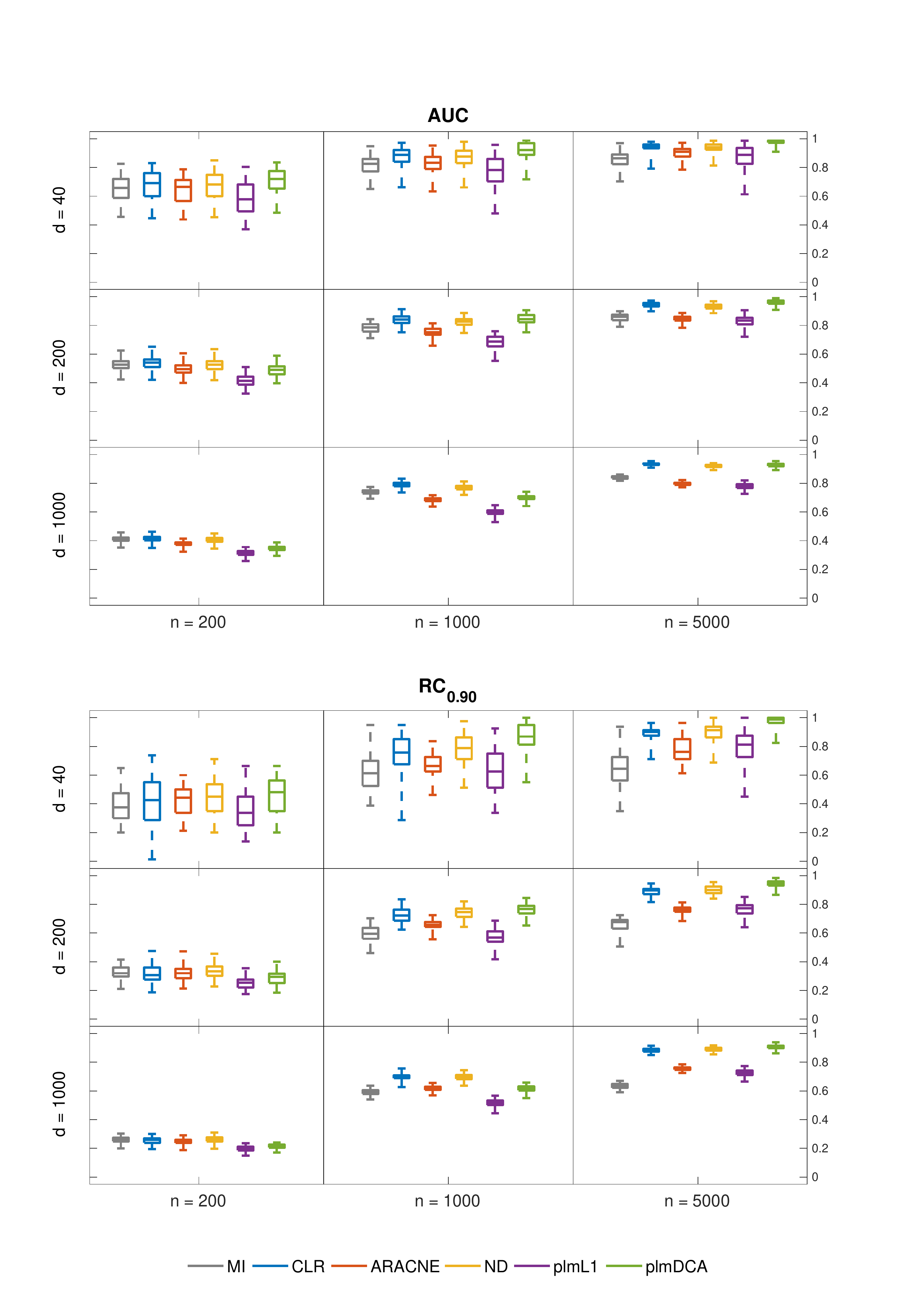}
\end{center}
\caption{Small-world network: Area under the precision-recall curve (AUC) and maximum recall for a precision of 0.90 ($RC_{0.90}$). \label{fig:res_sw}}
\end{figure}

The results of the experiments for the grid, scale-free and small-world model are summarized in Figures \ref{fig:res_grid}, \ref{fig:res_ba} and \ref{fig:res_sw}, respectively. The upper panels in the figures contain the AUC results and the lower panels contain the $RC_{0.90}$ results. For an overview of the average runtime of the methods, see Table \ref{tab:runtime}. We begin by discussing some more general observations before focusing in on the specific methods.

When comparing the overall accuracy on the different network types, the balanced grid network is the easiest to infer, closely followed by the small-world network, while the scale-free network is the most challenging. Other than that, the results are overall quite consistent across the different networks, however, there is one clear exception. ARACNE shows a drop in accuracy on the small-world network (Figure \ref{fig:res_sw}) in comparison to the other methods and networks (Figures \ref{fig:res_grid} and \ref{fig:res_ba}). The reason for this is that the small-world model is designed to generate clusters containing a lot of loops of length three (as confirmed by the clustering coefficient in Table \ref{tab:network_prop}). As explained in Section \ref{sec:aracne}, ARACNE is not able to recover loops of length three under a zero tolerance threshold making it particularly ill-suited for the small-world network. As suggested in \cite{Margolin06}, one could use a non-zero tolerance threshold allowing for some three-loops, however, it is not clear how to choose an appropriate non-zero threshold in practice. 

As expected, when comparing results across the rows and columns in the boxplot panels in Figures \ref{fig:res_grid}--\ref{fig:res_sw}, we see that the positive effect on accuracy given by a fivefold increase of the sample size $n$ is larger than the corresponding negative effect of a fivefold increase of the model size $d$. For example, the results found on the diagonal in the boxplot panels correspond to different settings for a fixed $n/d$-ratio of 5. When moving from the upper left corner to the lower right corner, the accuracy is increased.

When comparing the pairwise methods for the smallest sample size $n=200$, they all reached an accuracy comparable to plain MI. For larger samples, however, the post-processing step of the pairwise methods significantly improved the accuracy compared to plain MI (except for ARACNE on the small-world network). The need for filtering out indirect dependencies is particularly clear when examining $RC_{0.90}$. Except for the small-world network, we did not observe any major differences in accuracy between the pairwise methods. In terms of AUC, CLR was marginally more accurate than ARACNE and ND. In terms of $RC_{0.90}$, ARACNE (grid and scale-free network) and ND (all networks) were slightly more accurate than CLR. 

When comparing the pseudo-likelihood methods, the $L_2$-based plmDCA clearly outperformed the $L_1$-based plmL1 throughout this experiment in terms of both evaluation criteria. Overall, plmL1 had a surprisingly low accuracy in this experiment and was even outperformed by plain MI for all but the largest samples. Consequently, we focus on plmDCA in the remaining summary of the simulation results.

When comparing plmDCA and the pairwise methods, the results show that plmDCA requires a relatively large sample to gain an advantage over the pairwise methods through its global approach. Overall, plmDCA and the pairwise methods achieved a comparable accuracy in settings where the $n/d$-ratio was 5 (diagonal in boxplot panels). The pairwise methods achieved a higher accuracy in the small-sample settings where the $n/d$-ratio was 1/5 and 1 (below diagonal in boxplot panels), while plmDCA reached a higher accuracy in the large-sample settings where the $n/d$-ratio was 25 and 125 (above diagonal in boxplot panels). Consequently, in terms of accuracy, simple pairwise methods should be preferred for small samples, while plmDCA should be preferred for large samples. Due to the multifactorial nature of the structure learning problem, it is difficult to give a rule of thumb for what constitutes a large sample size in a more general sense. Still, purely based on these experiments, $n$ should be close to five times larger than the number of variables when $d$ is in the range of 40--1000. When scaling up even further, we would expect the required $n/d$-ratio to come down, however, we cannot verify this in practice simply due to computational reasons.

\section{Discussion \label{sec:discussion}}

We have empirically studied various methods for high-dimensional structure learning of binary pairwise Markov networks. We considered a collection of methods from the families of \emph{pseudo-likelihood methods}, developed in the fields of statistics and statistical mechanics, and \emph{pairwise methods}, developed in the field of computational biology. Since the methods have been developed in parallel in different fields, they have not (as far as we are aware) been compared against each other in a context similar to the one considered in this work. Conceptually, pseudo-likelihood methods and pairwise methods are very different. Pseudo-likelihood methods take a global approach, considering all variables simultaneously. In contrast, pairwise methods are based on local pairwise tests, which are post-processed in an attempt to adjust for the global aspect of the problem. Whereas the pseudo-likelihood methods enjoy nice theoretical properties such as statistical consistency, the main motivation behind pairwise methods is typically considered to be computational efficiency.

Although consistency is a desirable property in a structure learning method, it does not guarantee superior performance for finite-sized samples. The numerical results in this work showed that various pairwise methods can compete with, and even outperform, the more elaborate pseudo-likelihood methods, not only in terms of speed, but also in terms of accuracy on large synthetic networks when $n\leq d$, which is a common setting in many real-world structure learning problems. As score-based methods are generally thought to be more accurate than constraint-based methods, this finding is not only surprising, but also a valuable insight from a more practical perspective. In particular, as the size of the considered problems keeps increasing, it becomes ever more important to keep down the computational load of the learning algorithms. A recent example of an ultra-high-dimensional structure learning application (with $n<<d$) is genome-wide epistatic interaction discovery \cite{Skwark17, Puranen18}, where the networks contain tens of thousands or even up to a hundred thousand single nucleotide polymorphisms.

As part of this work, we also presented an sRBM-based Gibbs sampler, similar to the one proposed in \cite{Martens10}. To further scale up future studies to an ultra-high-dimensional setting, efficient sampling from a ground-truth model is a critical step. Although the sRBM sampler requires an additional step to sample the hidden variables, the main advantage of the sampler, in comparison to the standard sequential sampler, comes from the possibility of sampling the individual nodes in parallel. By exploiting massively parallel platforms, this could enable efficient sampling of ultra-high-dimensional sparse Markov networks over hundreds of thousands of variables.

In this work, we sidestepped the problem of determining a significance threshold for the interaction strength. In practical applications, one is typically only interested in edges that are deemed significant by some statistical criterion. Moreover, the computational efficiency of methods such as ARACNE depends on the possibility to remove edges with non-significant MI values before applying the DPI step. This is the reason behind the seemingly high runtimes of ARACNE compared to the other pairwise methods (see Table \ref{tab:runtime}). The perhaps most common general-purpose method for determining a significance threshold is to use permutation tests in which the values obtained for the original data are compared to values obtained for a corresponding permuted data set, where each column has been randomly rearranged in order to make the variables mutually independent. For example, permutation-based procedures have been used to obtain a significance threshold for mutual information in \cite{Butte00,Margolin06}. Alternatives approaches for determining significance in the model-based pseudo-likelihood methods are based on cross-validation, the extended Bayesian information criterion (eBIC) \cite{Barber15} and inverse finite-size scaling \cite{Xu18}. 

\section*{Acknowledgements}
This research was supported by the COIN Centre of Excellence, Academy of Finland grant no. 251170 (JP, YX, SP, MP, JC), KAKENHI no. 17H00764 (YK), and ERC grant no. 742158 (JC). The authors wish to acknowledge CSC -- IT Center for Science, Finland, for computational resources.

\section*{References}
\bibliographystyle{elsarticle-num}
\bibliography{sim_study}

\newpage
\appendix
\section{Implementation details \label{app:impl}}
The experiments were run in MATLAB using primarily existing and publicly available code packages. The methods were implemented and set up as follows:
\begin{itemize}
\item \emph{sRBM sampler}: the sRBM based Gibbs sampler was implemented in MATLAB, and is available for download at \url{https://bitbucket.org/jopensar/srbm-sampler/}. The burn-in and thinning were set to 2000 and 50 iterations, respectively.
\item \emph{Network generation:} the scale-free and small-world networks were generated using the CNM toolbox \cite{Alanis-Lobato14}. The parameters of the scale-free model were set to $m_0 = 3$ and $m = 2$. The parameters of the small-world model were set to $k = 4$ and $p = 0.25$. 
\item \emph{plmL1}: The $L_1$-regularization paths were calculated using the L1General MATLAB package \cite{Schmidt10,L1General}. The regularization weight for the bias parameter was set to 0 and the range of regularization weights for the interaction parameters was set according to 
\begin{equation*}
\lambda \in \{ \frac{k}{100}\lambda_{\max} \}_{k=1}^{100},
\end{equation*}
where $\lambda_{\max}$ was the minimum weight for which all interactions parameters were forced to zero (adapted from \cite{L1General}). Given a value on the regularization weight, the graph was constructed using the max criterion. By moving over the range of considered $\lambda$-values, from large to small, more edges were successively added according to their estimated importance. The regularization weight for which an edge was included thus serves as a measure of interaction strength, which was ultimately used to rank the edges. 

\item \emph{plmDCA}:  The authors' code package was used (\url{https://github.com/magnusekeberg/plmDCA}). The regularization weights were set according to the authors' recommendations, which are
\begin{equation*}
\begin{cases}
\lambda = 0.01\cdot n & \mbox{, if } n > 500 \\
\lambda = \left[ 0.1-(0.1-0.01)\dfrac{n}{500}\right] \cdot n & \mbox{, if } n \leq 500
\end{cases}
\end{equation*}
for the bias parameters and $\lambda/2$ for the interaction parameters, since they are regularized twice in the asymmetric plmDCA version. No sequence re-weighting was applied as the observations were (approximately) i.i.d. The reported results were obtained using average-product correction, since it gave marginally better results than without average-product correction.

\item \emph{MI}: A function for calculating the MI values according to the specifications in Section \ref{sec:mi} was implemented in MATLAB. 
\item \emph{CLR}: The CLR step was performed using the authors' code package (\url{http://m3d.mssm.edu/network_inference.html}) under default settings.
\item \emph{ARACNE}: The ARACNE post-filtering step was implemented in MATLAB. The tolerance threshold was set to zero, meaning that the weakest edge was always removed from a triplet. The mutual information of edges that were selected for removal was set to 0.
\item \emph{ND}: The ND step was performed using the authors' code package for symmetric networks (\url{http://compbio.mit.edu/nd/code/ND.m}) under default settings.
\end{itemize}
To obtain a complete ranking for the precision-recall curves for each method, all edges were sorted according to their estimated interaction strength. 

\newpage
\section{Runtime of methods\label{app:runtime}}
\setlength{\tabcolsep}{23pt}
\begin{table}[h]
\begin{center}
\small
\begin{tabular}{l l r r r}
\toprule
Method & $d$ & \multicolumn{3}{c}{Runtime}  \\
\cmidrule(lr){3-5}
& & $n=200$ & $n=1000$ & $n=5000$ \\
\midrule
& 40 & 0.06 & 0.02 & 0.07 \\ 
MI & 200 & 0.40 & 0.53 & 1.34 \\ 
& 1000 & 8.99 & 12.38 & 28.50 \\ 
\\
& 40 & 0.05 & 0.00 & 0.00 \\ 
CLR & 200 & 0.03 & 0.02 & 0.02 \\ 
& 1000 & 2.35 & 2.11 & 2.01 \\ 
\\
& 40 & 0.22 & 0.18 & 0.18 \\ 
ARACNE & 200 & 5.12 & 5.12 & 5.16 \\ 
& 1000 & 173.73 & 176.73 & 177.69 \\ 
\\
& 40 & 0.13 & 0.00 & 0.00 \\ 
ND & 200 & 0.06 & 0.03 & 0.03 \\ 
& 1000 & 2.39 & 2.44 & 2.53 \\ 
\\
& 40 & 17.07 & 18.28 & 37.18 \\ 
PlmL1 & 200 & 178.55 & 242.67 & 594.57 \\ 
& 1000 & 1729.01 & 8028.80 & 31062.33 \\ 
\\
& 40 & 0.62 & 1.62 & 6.42 \\ 
PlmDCA & 200 & 10.25 & 37.50 & 173.20 \\ 
& 1000 & 240.88 & 1286.27 & 8171.49 \\ 
\bottomrule 
\end{tabular}
\end{center}
\caption{Average runtime of methods (in seconds). The table shows the runtime of the postprocessing step of the pairwise methods. The total runtime of a pairwise method is obtained by adding the runtime of MI.\label{tab:runtime}}
\end{table}

\end{document}